\documentclass[conference]{IEEEtran}
\IEEEoverridecommandlockouts
\usepackage{cite}
\usepackage{amsmath,amssymb,amsfonts}
\usepackage{algorithmic}
\usepackage{graphicx}
\usepackage{textcomp}
\usepackage{url}
\usepackage{xcolor}
\def\BibTeX{{\rm B\kern-.05em{\sc i\kern-.025em b}\kern-.08em
    T\kern-.1667em\lower.7ex\hbox{E}\kern-.125emX}}
\begin{document}

\title{Rel-UNet: Reliable Tumor Segmentation via Uncertainty Quantification in nnU-Net}

\author{\IEEEauthorblockN{1\textsuperscript{st} Seyed Sina Ziaee*}
\IEEEauthorblockA{\textit{University of Calgary} \\
Calgary, Canada \\
0000-0003-3471-6560}
*Corresponding author
~\\
\and
\IEEEauthorblockN{2\textsuperscript{nd} Farhad Maleki}
\IEEEauthorblockA{\textit{University of Calgary} \\
Calgary, Canada \\
farhad.maleki1@ucalgary.ca}
~\\
\and
\IEEEauthorblockN{3\textsuperscript{rd} Katie Ovens}
\IEEEauthorblockA{\textit{University of Calgary} \\
Calgary, Canada \\
katie.ovens@ucalgary.ca}
~\\
}

\maketitle

\begin{abstract}
Accurate and reliable tumor segmentation is essential in medical imaging analysis for improving diagnosis, treatment planning, and monitoring. However, existing segmentation models often lack robust mechanisms for quantifying the uncertainty associated with their predictions, which is essential for informed clinical decision-making. This study presents a novel approach for uncertainty quantification in kidney tumor segmentation using deep learning, specifically leveraging multiple local minima during training. Our method generates uncertainty maps without modifying the original model architecture or requiring extensive computational resources. We evaluated our approach on the KiTS23 dataset, where our approach effectively identified ambiguous regions faster and with lower uncertainty scores in contrast to previous approaches. The generated uncertainty maps provide critical insights into model confidence, ultimately enhancing the reliability of the segmentation with the potential to support more accurate medical diagnoses. The computational efficiency and model-agnostic design of the proposed approach allows adaptation without architectural changes, enabling use across various segmentation models.
\end{abstract}

\begin{IEEEkeywords}
Uncertainty Quantification, Tumor Segmentation, nnUNet, Medical Image Segmentation
\end{IEEEkeywords}

\section{Introduction}
Medical imaging plays a crucial role in the diagnosis, treatment planning, and monitoring of cancer. Accurate and reliable tumor segmentation improves patient outcomes and enables precision medicine. Despite advancements in imaging technologies, accurate tumor segmentation remains challenging due to the heterogeneous nature of tumors, variations in imaging modalities, and the presence of noise and artifacts 
\citenum{saman2019survey}. These challenges can lead to significant variability in segmentation results, which impacts clinical decision-making and patient care \cite{zou2023review}. To ensure the reliability of automated segmentation models in clinical practice, it is crucial to predict the tumor boundaries and quantify the uncertainty associated with these predictions. 

Uncertainty quantification provides insights into model confidence, identifies potential errors, and helps in risk assessment, supporting more informed clinical decisions. While several deep learning-based methods have achieved promising results in tumor segmentation with uncertainty aware approaches, they often need complex mechanisms for uncertainty quantification. Some of the main downsides of the existing approaches include providing overly conservative estimates, being computationally expensive, or changing the model architecture to calculate uncertainty \cite{heller2021state, yogananda2020fully, roy2022uncertainty, dorta2018structured}.

In this paper, we present a new approach in estimating uncertainty in image segmentation, and demonstrate its application in kidney tumor segmentation. Our method leverages multiple checkpoints from the local minima obtained during the training process of the segmentation model to generate robust uncertainty maps in one training and inference process. This approach provides critical insights into model confidence and improves segmentation accuracy without changing the model architecture or using computationally expensive resources. We evaluated our approach on the KiTS23 dataset \cite{heller2023kits21}, demonstrating its effectiveness in identifying ambiguous regions and out-of-distribution data. The generated uncertainty maps could enhance clinical decision-making by highlighting areas of low confidence, ultimately aiding in more accurate and reliable medical diagnoses. The implementation of this work is available at \cite{code}.

\section{Literature Review}

Techniques that have been introduced to estimate uncertainty in medical segmentation tasks can be categorized into deterministic single networks, Bayesian Neural Networks (BNNs), ensemble methods, and test-time data augmentation approaches \cite{zou2023review}. 

In deterministic single networks, uncertainty is inferred from the output of the network, where it is assumed that the model is deterministic, and the uncertainty can be estimated based on one single forward pass.
Holder et al. \cite{holder2021efficient} introduced Uncertainty Distillation, where a student network learns from a teacher network’s output, including its uncertainty, to improve confidence estimation. Franchi et al. \cite{franchi2021one} presented OVNNI, a system using "One-vs-All" and "All-vs-All" networks to handle uncertainty, collaborating to detect out-of-distribution data through confidence scores.

One of the main downsides of deterministic single networks is their limitation in scenarios with limited data, leading to high-confidence predictions even in areas where the model should be uncertain. Bayesian Neural Networks (BNNs) address this limitation by treating the network's weights as probability distributions. This allows the network to learn a distribution of possible predictions instead of a deterministic prediction obtained with the point estimate values of the network parameters, inherently capturing the uncertainty of the model. Salahuddin et al. \cite{salahuddin2023leveraging} proposed a modification to a cascaded nnU-Net framework \cite{nnunet} for segmenting kidneys, tumors, and cysts in CT scans. This approach leverages uncertainty estimation through Monte Carlo Dropout to identify potentially ambiguous structures and improve segmentation accuracy, particularly for tumors and cysts. Baumgartner et al. \cite{baumgartner2019phiseg} proposed a method that leverages a hierarchical probabilistic model to capture the uncertainty in medical image segmentation. Unlike traditional deterministic models, PHISeg estimates a distribution over possible segmentations by learning a conditional variational autoencoder. This allows the model to produce a variety of plausible segmentations, reflecting both model and data uncertainty.
Zhao et al. \cite{zhao2022efficient} introduced an efficient approach for posterior sampling of weight space to estimate Bayesian uncertainty. Their approach maintains the original segmentation model architecture without requiring a variational framework, thereby preserving the performance of nnU-Net. Additionally, they boost the segmentation performance over the original nnU-Net via marginalizing multi-modal posterior models. They applied their method to the public ACDC \cite{SDV21} and M\&M \cite{campello2021multi} datasets of cardiac MRI.

In contrast to BNNs, deep ensemble models, train multiple independent models or data and average their predictions to capture uncertainty. They are easier to implement and scale but less theoretically rooted in probability. The final prediction is often an aggregation of the individual predictions, leveraging the disagreement among the ensemble members. 
Causey et al. \cite{causey2021ensemble} proposed to train multiple variations of the U-Net architecture and data augmentation, where each U-Net votes on kidney and tumor segmentation in new CT images, and their combined predictions yield the final segmentation. 
Khalili et al. \cite{khalili2024uncertainty} proposed the Uncertainty-Guided Annotation (UGA) framework, enabling AI to convey pixel-level uncertainties generated from 5-fold cross-validation trained models to clinicians for improved model performance. UGA enhanced the Dice coefficient on the Camelyon dataset through clinician feedback.

Another approach of uncertainty estimation is Test-Time Data Augmentation (TTA), where it focuses on creating variations of the input data through techniques like random crops, rotations, or adding noise. Methods in this area are easier to implement as no change to the original segmentation model is needed. The model's predictions for these augmented versions are used for uncertainty estimation, where high variability in the network's predictions for these variations suggests the model is sensitive to slight changes in the input and, hence, less certain about the prediction. 
Wang et al. \cite{wang2019aleatoric} explored the application of TTA with uncertainty estimation for deep learning-based medical image segmentation in a broader context. They compared TTA-based uncertainty to model-based uncertainty estimation (limitations of the model itself) and demonstrated the benefits of TTA for improving segmentation performance and reducing overconfident incorrect predictions.

One of the significant downsides of many of the aforementioned approaches, particularly the Monte Carlo dropout and other Bayesian Neural Networks, is that they change the model architecture, which leads to decreasing the performance of the models that have been fine-tuned on specific tasks. Even when using TTA-based approaches, where the model architecture is not changed, the predictions are often biased by the applied augmentations, leading to inconsistencies. Our approach for uncertainty quantification is independent of the model architecture and consistent, where we sample multiple checkpoints from the local minima in the training process. These checkpoints generate uncertainty maps that indicate the regions in which the model is uncertain about in the segmentation.

\section{Dataset and Methodology}

\begin{figure*}[t]
    \centering
    \includegraphics[width=1.0\linewidth]{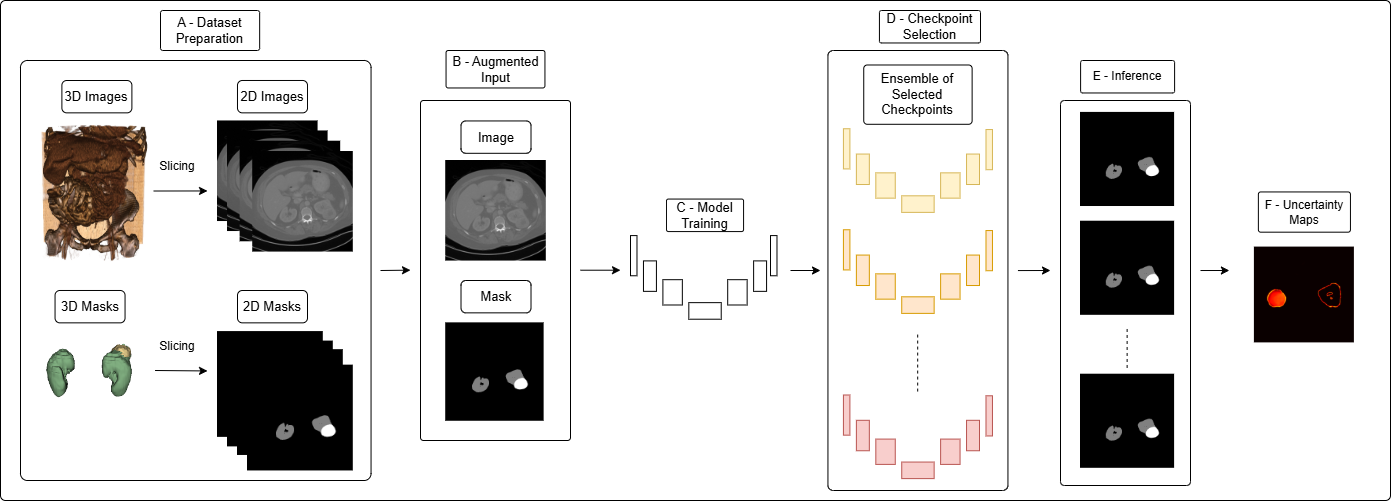}
    \caption{Overview of the methodology. In A, dataset is prepared for model training and inference. In B, the augmented input is fed into the training model in C. In D, after training is finished, an ensemble of checkpoints are selected from the saved checkpoints. In E, ensemble of the models are used to generate predictions based on the test data input, and in F, uncertainty maps are generated from the perturbations in ensemble of predictions.}
    \label{fig:pipeline}
\end{figure*}

The steps of our method are outlined in Figure~\ref{fig:pipeline} and described in the following subsections.

The steps of our method are described in the following subsections.

\subsection{Dataset}
We evaluated our method on the Kidney Tumor Segmentation Challenge (KiTS23) dataset \cite{heller2023kits21}, a recent and comprehensive dataset of cross-sectional medical images for kidney and kidney tumor segmentation. The dataset is composed of 489 cases (patients). Each case contains labels for kidney, tumor, and cyst, with pixel values of 1, 2, and 3 respectively.

\subsection{Pre-processing}
\begin{itemize}
    \item \textbf{Dataset Splitting}: We split the dataset into 400 cases for training and validation, and the rest for testing. Using 5-fold cross-validation, 320 cases are used for training, and 80 cases are used for validation.
    \item \textbf{Resampling}: Different cases include different voxel spacing 
    , and to ensure common voxel spacing and reduce variability in spatial resolutions, we perform resampling.
    \item \textbf{Intensity Normalization}: Intensity values are normalized by subtracting the mean and dividing by the standard deviation, done on a per-case basis, normalizing each 3D volume independently to improve model training.
    \item \textbf{Slicing}: For each 3D CT scan volume, we perform 2D axial slicing to generate 2D images. After inference and prediction of the model, they are stacked up on each other to reconstruct the images similar to the original 3D volumes. 
    \item \textbf{Data Augmentation}: We apply several data augmentation techniques to increase the diversity of the training data, including rotations, scaling, mirroring, gamma correction, flipping, and rotating.
    
\end{itemize}

\begin{figure}[t]
    \centering
    \includegraphics[width=1.0\linewidth]{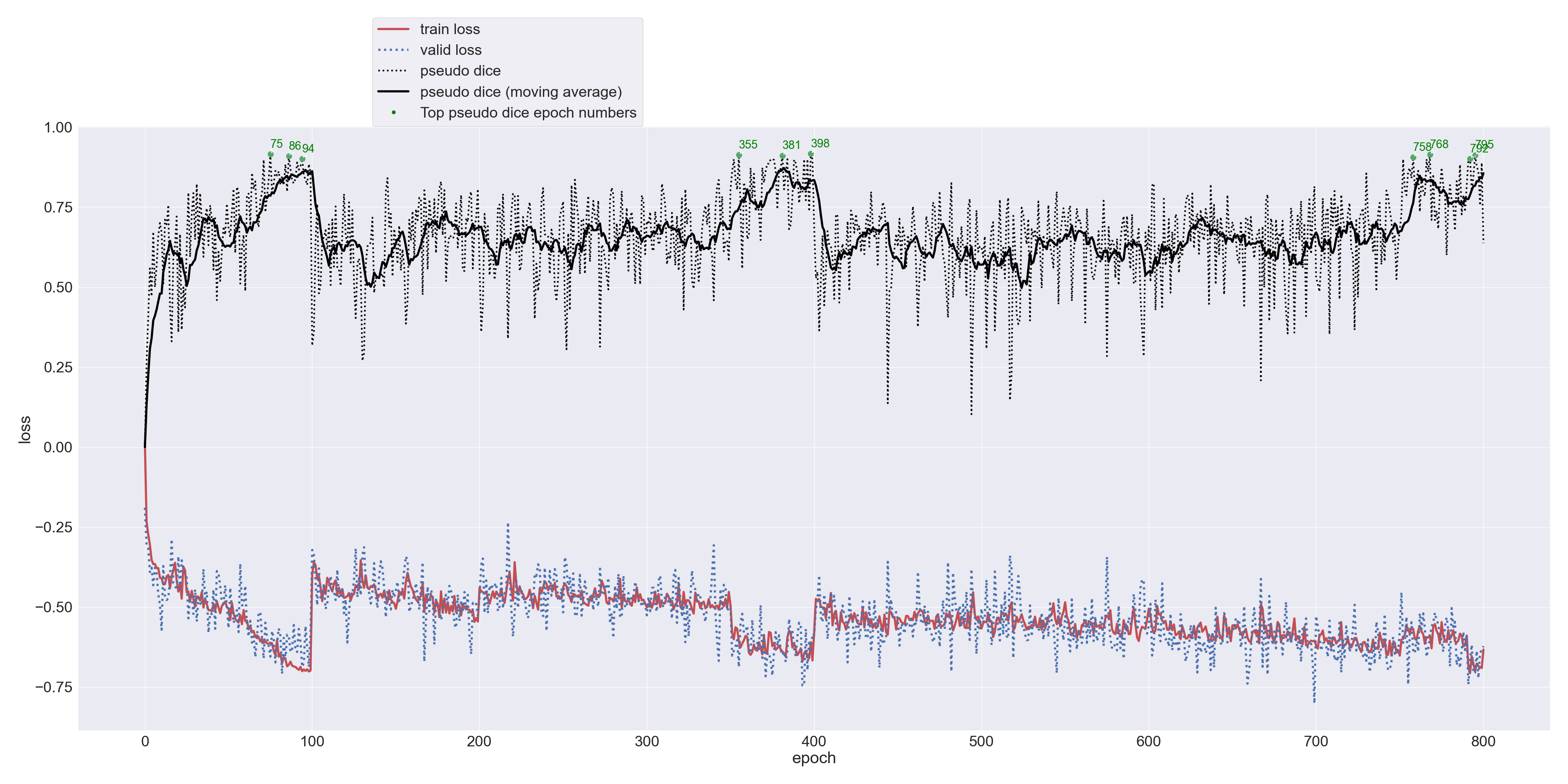}
    \caption{a) The learning rate value over the training progress with the SGDR learning rate scheduler. b) Training progress over 800 epochs and 4 cycles. Green numbers indicate epoch numbers, where the local minimum is happening (highest validation Dice score) in the training progress. These are around epochs 100, 200, 400, and 800, where the top three checkpoints around each local minimum is selected for uncertainty map generation. Note that model training progress does not show any peak around epoch 200. Thus, no checkpoint is selected around this epoch.}
    \label{fig:training-progress}
\end{figure}

\subsection{Segmentation Model (Training and Inference)}
Our methodology is built upon nnU-Net \cite{nnunet}, a deep learning framework specifically designed for medical image segmentation tasks. It tackles the challenge of adapting segmentation pipelines to new datasets by automatically extracting key dataset characteristics and using them to configure a U-Net-based architecture with optimal parameters. This eliminates the need for manual configuration and improves generalizability to unseen datasets. We specifically used nnU-Net ResEnc Large as the segmentation model. We performed a 5-fold cross-validation to have a better evaluation. The nnU-Net framework was used for both training and inference of the model. The original nnU-Net method used polynomial learning rate decay, where the learning rate decreases to nearly zero in the last epochs of the training phase. However, we updated the learning rate scheduler, which is adapted using Stochastic Gradient Descent with Warm Restarts (SGDR). This method allows the model to escape local minima by periodically restarting the learning rate to a higher value, creating multiple local minima before the restart happens in the training progress.

As shown in Figure~\ref{fig:training-progress}, the model has three peaks based on the pseudo Dice as a metric. Then, the learning rate scheduler modifies the learning rate. In this way, the last peak of the model has better performance in segmentation, and we sample posterior weights from the peaks of this training progress. In other words, we take the checkpoints from the peak points of the training process. Next, we use the ensemble of these checkpoints to generate segmentation and uncertainty maps. These peaks are the local minima. 
The goal is to use the model's local minima obtained by the previous peaks (epochs 100 and 400 in our case) and the optimal minima in the last peak to have an ensemble of models for uncertainty quantification. This approach is independent of the model architecture and can be used with any other model or framework for quantifying uncertainty. The approach is computationally efficient because it creates an ensemble effect at inference time without requiring the training of multiple models.
The learning rate formula used in this methodology is as follows:

\begin{equation}
T_{i} = T_0 \cdot \eta^{i}
\end{equation}
\begin{equation}
\text{lr}(t) = \frac{lr_{min}}{2} \left( \cos \left( \frac{\pi \cdot \text{mod}(t, T_{i})}{T_{i}} \right) + 1 \right) + lr_{min}
\end{equation}

In the given formula, SGDR adjusts the learning rate in a cyclical manner, where each cycle's length is determined by $T_i = T_0 \cdot \eta^i$, with $T_0$ as the initial cycle length and $\eta$ as the factor by which the cycle length increases after each restart. Within each cycle, the learning rate $\text{lr}(t)$ follows a cosine annealing schedule, starting from a higher value and gradually decreasing to a lower bound, $lr_{min}$. 
Using the cosine function ensures a smooth decrease in the learning rate within each cycle, and the periodic warm restarts at the end of each cycle allow the model to escape local minima by resetting the learning rate and improving convergence while exploring different regions of the loss landscape. Figure~\ref{fig:training-progress} demonstrates the training progress.

\subsection{Checkpoint Selection}
Three checkpoints are selected from the peaks of each cycle, represented by green numbers as indicated in Figure~\ref{fig:training-progress}. Predictions are made with each of the sampled checkpoints. The ensemble of these predictions creates uncertainty maps as models tend to predict different segmentations, especially around the borders of kidneys, tumors, and cysts.

\subsection{Bayesian Inference}
 This step is done for four types of classes, which are background, kidney, tumor, and cyst. We generate four probability maps and estimate an average of individual checkpoint probability, from the test output \(y_{test}\), given the test input \(x_{test}\) and dataset \(D\) using the following formulas:

\begin{equation}
p(y_{\text{test}}|x_{\text{test}}, D) \approx \frac{1}{n} \sum_{i=1}^{n} p(y_{\text{test}}|x_{\text{test}}, w_{t_i}) , \text{for } w_{t_i} \in W 
\end{equation}



The ensemble prediction is approximated by averaging the predictions \(p\) from multiple checkpoints, where \( n \) is the number of checkpoints. \( w_{t_i} \) are the model parameters at checkpoint \( t_i \). The set of all model parameters \( w_{t_i} \) is defined as \( W \).

\subsection{Uncertainty Map Generation}
In order to generate uncertainty maps, we compute class-specific entropy for kidneys, tumors, and cysts. The process is as follows: 

\subsubsection{Entropy Calculation}
We calculated the entropy of each pixel using the following formula:

\begin{equation}
\label{eq:pixel_entropy}
H(y_{\text{test}}^{ij}) = - \sum_{k=1}^{c} p(y_{\text{test}}^{ij} = k | x_{\text{test}}, D) \log_2 p(y_{\text{test}}^{ij} = k | x_{\text{test}}, D)
\end{equation}

In Equation~\ref{eq:pixel_entropy}, \( H(y_{\text{test}}^{ij}) \) denotes the entropy of the predicted output for pixel in row i and column j, \( p(y_{\text{test}}^{ij} = k | x_{\text{test}}, D) \) is the probability that the predicted output \( y_{\text{test}}^{ij} \) is equal to class \( k \) given the test input \( x_{\text{test}} \) and the dataset \( D \). \( \sum_{k=1}^{c} \) indicates that the summation is performed over all \( c \) classes. The entropy is defined as the sum of the product of the probability of each class and the logarithm of that probability.\\

\subsubsection{Normalization}

Due to the variation in segmentation areas across samples and classes, more than the regular mean approach for quantifying entropy is needed to achieve a single uncertainty score per image. We applied normalization by generating a dilated version of each image, subtracting it from the original image, and isolating only the segmentation contours. Dilation of the binary image effectively enlarges the segmented regions by expanding the boundaries. This dilation helps isolate the segmentation contours more effectively by emphasizing the borders between segmented regions and the background. This approach provided a normalized sum of entropy values for each image. The formulas for computing the normalized entropy are as follows:

\begin{equation}
\label{eq:uncert_score}
p_{\text{class}}(y_{\text{test}} | x_{\text{test}}, D) \approx \frac{1}{n} \sum_{i=1}^{n} p_{\text{class}}(y_{\text{test}} | x_{\text{test}}, w_{ti}) 
\end{equation}

Equation~\ref{eq:uncert_score} represents a process for calculating an uncertainty score from image segmentation outputs. It approximates the mean image of foreground probabilities for a given class, where the threshold is set to 0.5 to create a binary image \( I \). 

\begin{equation}
\label{eq:class_entropy}
H_{\text{total}} = \sum_{i} \sum_{j} H(y_{\text{test}})_{ij} 
\end{equation}

Equation~\ref{eq:class_entropy} \( H_{\text{total}} \) is the sum of entropy values across all pixels in the test image, reflecting the total uncertainty. 

\begin{equation}
\label{eq:dilate}
I^* = I \oplus A; 
\end{equation}

Equation~\ref{eq:dilate} describes \( I^* \) as the result of dilating the binary image \( I \) using a structuring element \( A \). Dilation is a morphological operation used to grow or expand the boundaries of regions of foreground pixels (in our case, it is pixel values of 1, 2, and 3 for kidney, tumor, and cyst, respectively) in a binary image. The structuring element (or kernel) defines the neighborhood where the dilation operation is performed.

\begin{equation}
\label{eq:normalize}
I_{\text{normalization}} = I - I^*
\end{equation}

Equation~\ref{eq:normalize} normalizes the binary image by subtracting the dilated image from the original binary image, yielding \( I_{\text{normalization}} \).

\begin{equation}
\label{eq:uncert_score_total}
\text{Uncertainty score} = \frac{H_{\text{total}}}{\sum_{i,j} I_{\text{normalization}}}
\end{equation}

Equation~\ref{eq:uncert_score_total} calculates the uncertainty score by dividing the total entropy \( H_{\text{total}} \) by the sum of the normalized image values \( I_{\text{normalization}} \). 

At this stage, we have obtained the entropy values and the inference results for each 2D slice within the dataset. By stacking these 2D slices, we reconstruct the uncertainty maps and segmentation masks similar to the original 3D images.

\begin{figure}[t]
    \centering
    \includegraphics[width=1.0\linewidth]{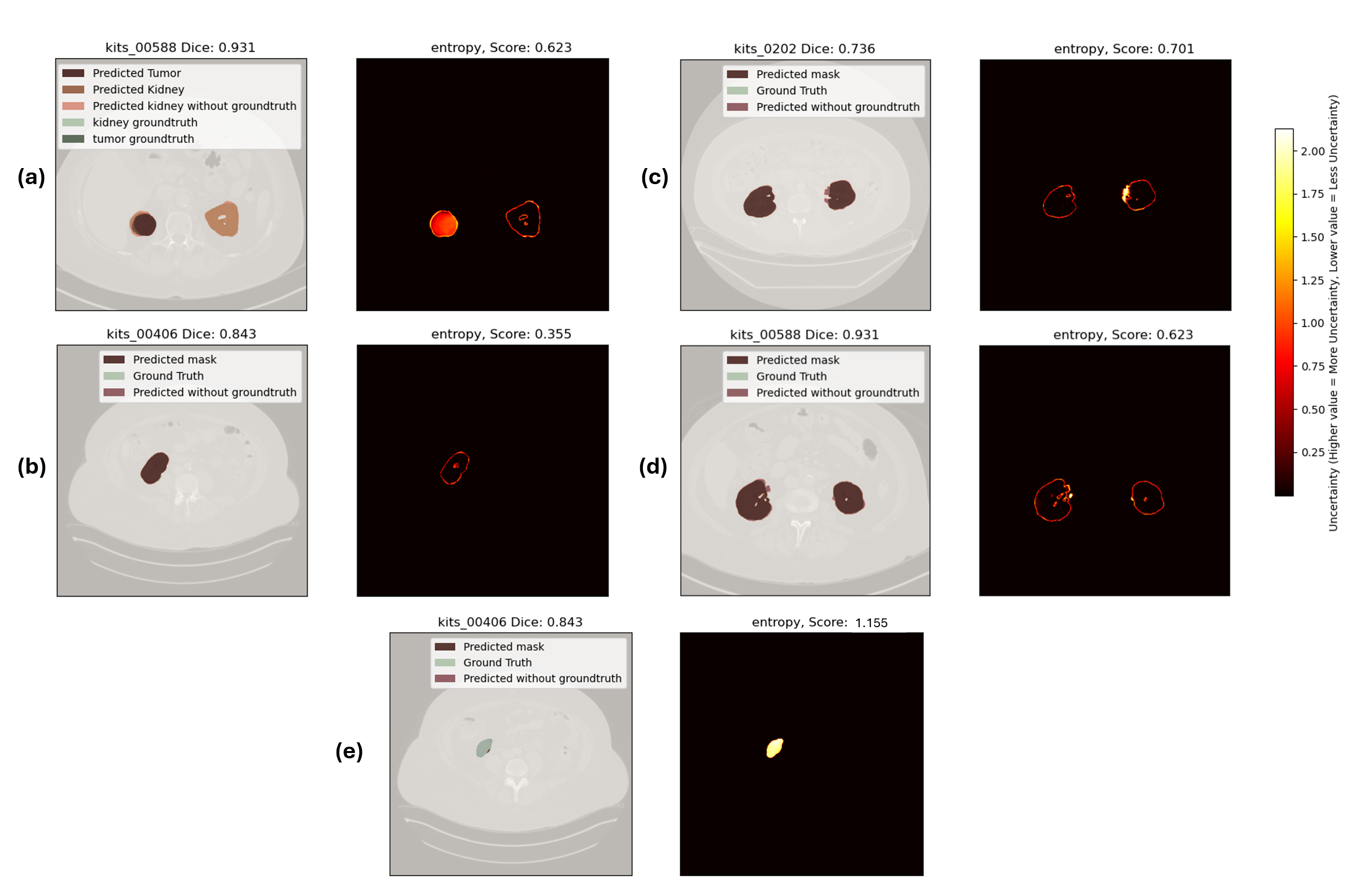}
    \caption{Samples of the results  comparing the segmentation performance with uncertainty maps. Lighter pixels indicate higher uncertainty and darker pixels indicate cases where the checkpoint models have the same inference results. In the labels of the images, "without ground truth" indicates pixels that the ground truth does not consider that part as a mask (kidney, tumor, or cyst), but the model has segmented that pixel. The Dice scores and uncertainty scores are for the whole 3D images and not the 2D slices in this figure alone.}
    \label{fig:good_results}
\end{figure}

\section{Results}
Comparison of select test samples based on their uncertainty maps and the segmentation performance is available in Figure~\ref{fig:good_results}. As shown in this figure, uncertainty maps help to complete some parts of the segmentation prediction, where the model does not show good performance. 

\begin{figure}[t]
    \centering
    \includegraphics[width=1.0\linewidth]{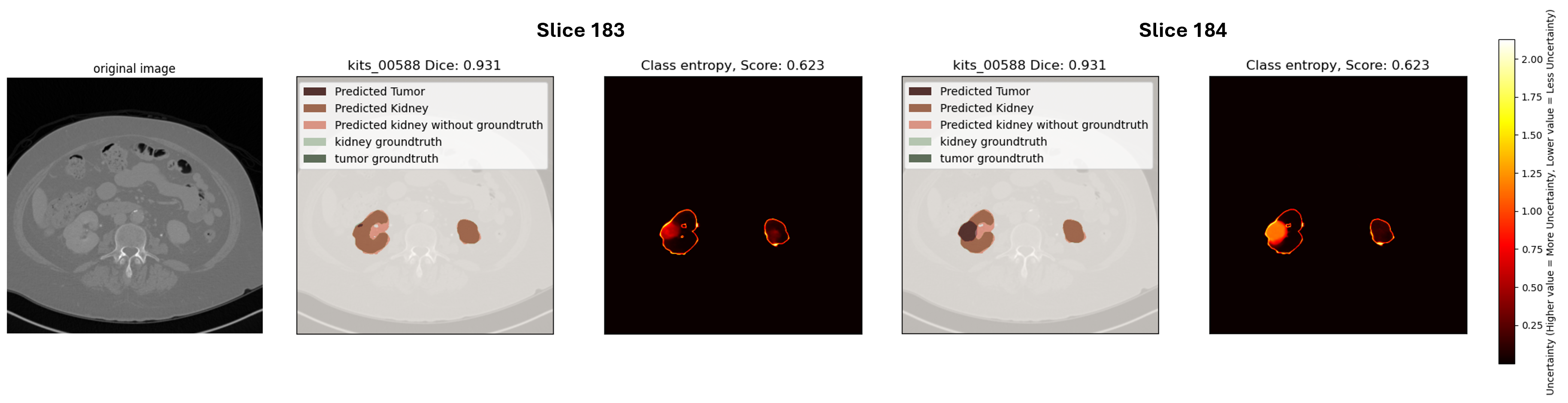}
    \caption{Uncertainty maps that show tumor presence in regions where both the segmentation model and ground truth are incomplete, as illustrated in consecutive slices 183 and 184. Slice 184 indicates a kidney slice where a tumor exists, and both the segmentation and the uncertainty map cover it. However, slice 183 (only one slice before 184) has no tumor in its ground truth, and the model's segmentation result shows a similar pattern. However, the uncertainty map demonstrates a tumor portion in this slice. So, while this slice has a tumor, neither the ground truth nor the segmentation map is showing it. As we have this evidence in exactly the next slice, we are sure that the uncertainty map is doing better in finding this slice as an unhealthy kidney with a tumor (Approved by a internal medicine specialist).}
    \label{fig:interesting_result}
\end{figure}

In some cases, the ground truth itself is not completely right, and with the help of the uncertainty maps, we find these areas. Figure~\ref{fig:interesting_result} demonstrates one of these cases.


\begin{table}[t]
\caption{Comparison of Expected Calibration Error (ECE) of our model with previous state-of-the-art uncertainty quantification models on KiTS23 dataset. Lower ECE indicates improved calibration performance. As shown, Rel-UNet, HMC \cite{zhao2022efficient}, and PHiSeg \cite{baumgartner2019phiseg} are among the top performers with lower ECE values. 5-fold \cite{khalili2024uncertainty}, Deep Ensemble \cite{causey2021ensemble}, and Monte Carlo \cite{monteiro2020stochastic} are next in line.}
    \centering
    \setlength{\tabcolsep}{8pt} %
    \begin{tabular}{c | c }
         Method & ECE (\%) \\
         \hline 
         5-fold & 2.89 \\
         PHiSeg & 2.52\\
         Monte Carlo & 3.44 \\
         Deep Ensemble & 2.70 \\
         HMC & 1.36 \\
         \textbf{Rel-UNet } & \textbf{1.11} \\
    \end{tabular}
    \label{tab:ece_comparison}
\end{table}

\begin{figure*}[ht]
    \centering
    \includegraphics[width=1.0\linewidth]{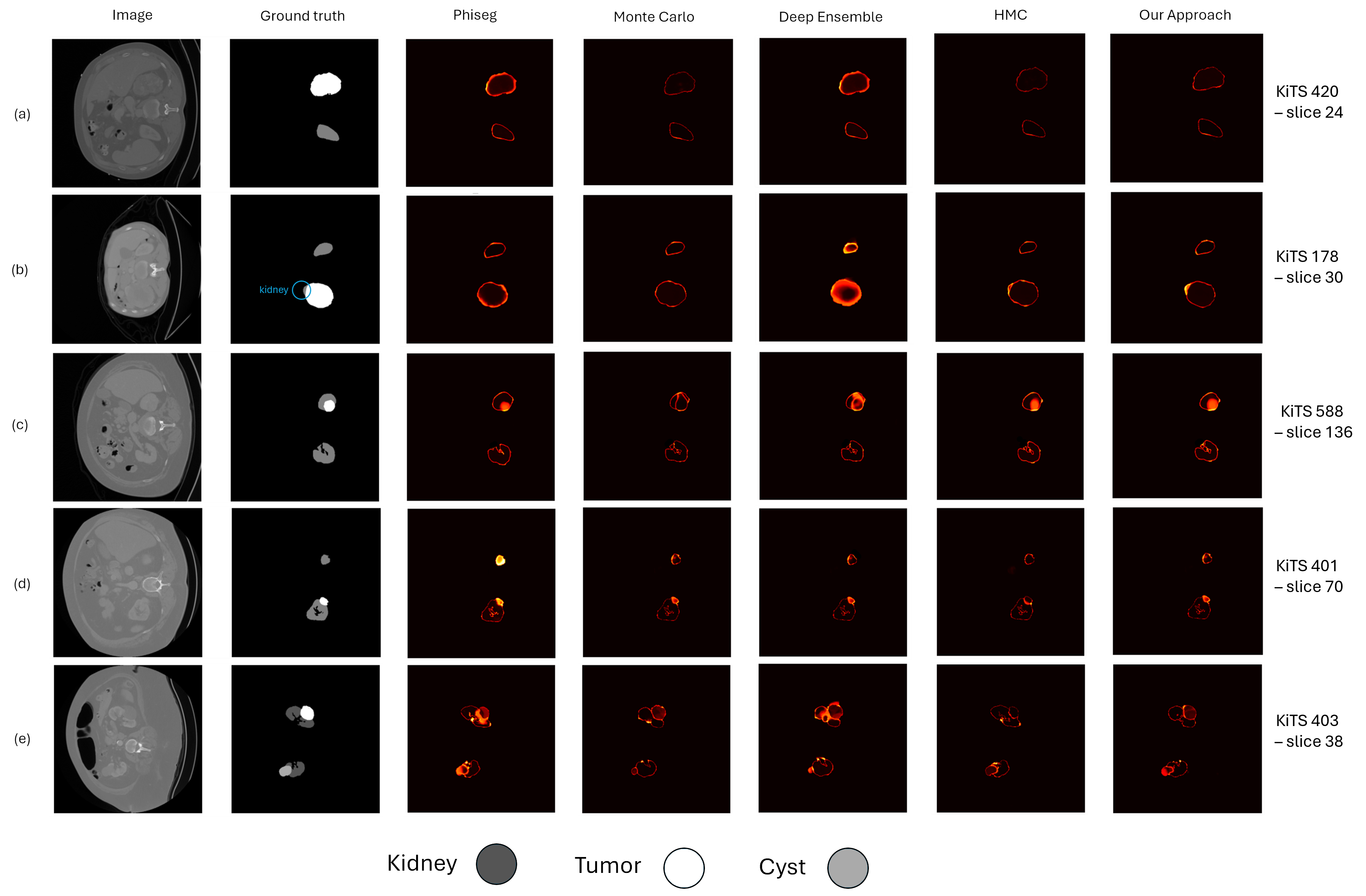}
    \caption{Comparison of our approach with previous state-of-the-art models in uncertainty estimation in uncertainty maps. Cases are KITS 420 (slice 24), KITS 178 (slice 30), KITS 588 (slice 136), KITS 401 (slice 70), and KITS 403 (slice 38), respectively.}
    \label{fig:comparison}
\end{figure*}

To quantitatively compare the results of our model on the test dataset with previous state-of-the-art models, we used Expected Calibration Error (ECE) that is available in Table~\ref{tab:ece_comparison}. ECE is a metric used to measure the difference between predicted probabilities and actual outcomes to assess the calibration of a model's confidence. It is computed by partitioning predictions into bins based on confidence scores and then calculating the weighted average of the absolute difference between accuracy and confidence within each bin. A lower ECE indicates that the model's predicted probabilities are well-calibrated to the true likelihood of segmentations. 

\begin{figure}[t]
\flushright
    \centering
    \includegraphics[width=1.0\linewidth]{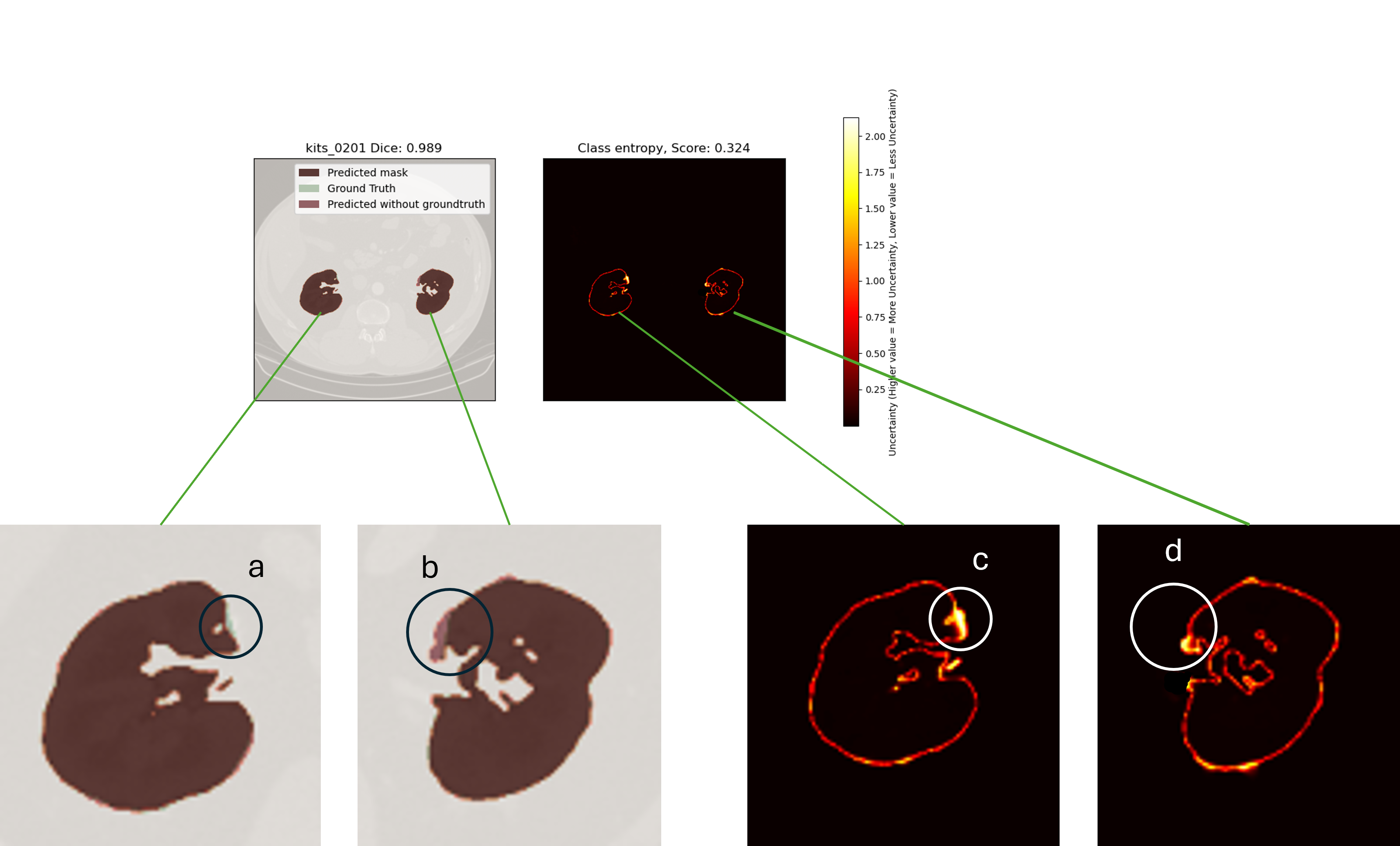}
    \caption{Example image where the ground truth does not include all portions of the kidney, yet the segmentation model captures additional areas. Circles a and c indicate the small portion of the kidney (green color) that the model could not predict and is highly uncertain about, which results are reflected in the uncertainty map. Circles b and d indicate portions of the kidney that are not available in the ground truth but are, in fact, a kidney portion. The segmentation model has found that part is the kidney, and the uncertainty is also high in this area (Approved by a specialist physician).}
    \label{fig:bad_segmentation}
\end{figure}

In addition, we qualitatively compared the result of the uncertainty maps generated by our model with previous state-of-the-art models in uncertainty quantification. Figure~\ref{fig:comparison} indicates the uncertainty maps that are generated for this comparison. 

In summary, our proposed method consistently outperformed baseline models in uncertainty estimation, with a 15\% reduction in Expected Calibration Error (ECE) and a comparable Dice score in tumor segmentation.

\section{Discussion}
In this study, our primary objectives were to generate uncertainty maps, calculate the Expected Calibration Error, and evaluate the reliability of the segmentation model, with segmentation accuracy as a secondary focus for evaluation purposes. We found that our proposed method outperformed a selection of state-of-the-art methods utilizing uncertainty to try to improve segmentation accuracy (shown in Table\ref{tab:ece_comparison}). 

Our evaluation also highlighted that the uncertainty maps generated by our method can detect areas missed by the segmentation model and ground truth, offering an additional evaluation that could improve diagnostic reliability. Figure~\ref{fig:good_results} shows some of the test data samples to compare the segmentation and uncertainty maps. Cases a, b, c, and d represent segmentations where the uncertainty map matches the predicted segmentation. In contrast, in some cases, the performance of the segmentation models is not good, but the uncertainty map has detected those parts as a potential kidney, tumor, or cyst portion. Case e indicates a case where the model segmentation has not detected the ground truth, the uncertainty is high and the uncertainty map has covered that area.

In addition, our evaluation highlights areas that the ground truth mask and the segmentation model match, but they are both wrong. In these cases, the labeling is wrong and the segmentation model is reflecting the same error in its prediction. 
Figure~\ref{fig:interesting_result} demonstrates a case where the segmentation model seems to match the ground truth, but both the ground truth and the segmentation model are wrong (case 183). This problem comes from the original annotation that the uncertainty map found, and it shows a tumor portion in the picture (in the left kidney).

In some cases, the ground truth is incomplete, and the segmentation model performs better than the ground truth. In Figure~\ref{fig:bad_segmentation}, in circle b, the ground truth is not considering the light red color as a kidney, but it is indeed a kidney portion, and we see the segmentation model has considered that portion as a kidney (Approved by a internal medicine specialist). This phenomenon also happens in case c in Figure~\ref{fig:good_results} on the right kidney. In addition, the model on circle a in Figure~\ref{fig:bad_segmentation}, has not detected a segment of the kidney, and the uncertainty in that segment is high in circle c. 

In Figure~\ref{fig:comparison}, cases (a) and (d) indicate a successful uncertainty map generation, where all approaches generate similar patterns for segmenting the kidney and the tumor available in the picture. Case (b) indicates a healthy kidney and an unhealthy kidney with a tumor, and the blue circle shows the kidney portion of the unhealthy kidney. In this case, Phiseg and Monte Carlo do not show the healthy portion of the unhealthy kidney, whereas deep ensemble, HMC, and our approach show that portion in their uncertainty map. In case (c), Monte Carlo and Deep Ensemble cannot find the tumor's position in the image, whereas Phiseg, HMC, and our approach do. Case (e) shows a relatively complex example where both tumor and cyst exist in the image, and previous approaches cannot fully locate the regions of the tumor, the cyst, and the healthy portions of the kidney accurately, while our method is locating these areas more precisely. 

Our approach of leveraging multiple local minima for uncertainty quantification presents a significant advancement over traditional methods, offering a computationally efficient solution without compromising segmentation accuracy. This could have broad implications for integrating uncertainty quantification into clinical workflows. 

While our study has provided valuable insights into the utilization of uncertainty maps to increase the level of trust and reliance in tumor and cyst segmentation models, several areas remain unexplored and present opportunities for further research.
First, Rel-UNet starts uncertainty estimation only after the training process is finished, which prevents the model from using the uncertainty estimation while the model is being trained. Second, in order to increase the usability of our model, we will incorporate it into other datasets and tumor segmentation tasks. Third, we will integrate our framework into other image segmentation models to enhance the reliability of their methodology.

\section{Conclusion}
We proposed a novel model-agnostic and efficient approach to quantifying uncertainty, designed to be applicable across various segmentation tasks and models. We showcased the utility of this method for uncertainty quantification in kidney tumor and cyst segmentation. The generated uncertainty maps resulted in enhancing segmentation accuracy and provided critical insights into model confidence. We demonstrated that these uncertainty maps improve segmentation accuracy and aid in identifying ambiguous regions, which is crucial for supporting more informed clinical decision-making. In future work, we aim to extend the applicability of this method to other segmentation tasks and medical datasets, further exploring how uncertainty estimation can enhance the performance and reliability of models in various medical contexts.


@inproceedings{jungo2019assessing,
  title={Assessing reliability and challenges of uncertainty estimations for medical image segmentation},
  author={Jungo, Alain and Reyes, Mauricio},
  booktitle={Medical Image Computing and Computer Assisted Intervention--MICCAI 2019: 22nd International Conference, Shenzhen, China, October 13--17, 2019, Proceedings, Part II 22},
  pages={48--56},
  year={2019},
  organization={Springer}
}

@article{franchi2021one,
  title={One versus all for deep neural network for uncertainty ({OVNNI}) quantification},
  author={Franchi, Gianni and Bursuc, Andrei and Aldea, Emanuel and Dubuisson, S{\'e}verine and Bloch, Isabelle},
  journal={IEEE Access},
  volume={10},
  pages={7300--7312},
  year={2021},
  publisher={IEEE}
}

@inproceedings{roy2022uncertainty,
  title={Uncertainty-guided source-free domain adaptation},
  author={Roy, Subhankar and Trapp, Martin and Pilzer, Andrea and Kannala, Juho and Sebe, Nicu and Ricci, Elisa and Solin, Arno},
  booktitle={European Conference on Computer Vision},
  pages={537--555},
  year={2022},
  organization={Springer}
}

@inproceedings{dorta2018structured,
  title={Structured uncertainty prediction networks},
  author={Dorta, Garoe and Vicente, Sara and Agapito, Lourdes and Campbell, Neill DF and Simpson, Ivor},
  booktitle={Proceedings of the IEEE Conference on Computer Vision and Pattern Recognition},
  pages={5477--5485},
  year={2018}
}

@article{zou2023review,
  title={A review of uncertainty estimation and its application in medical imaging},
  author={Zou, Ke and Chen, Zhihao and Yuan, Xuedong and Shen, Xiaojing and Wang, Meng and Fu, Huazhu},
  journal={Meta-Radiology},
  pages={100003},
  year={2023},
  publisher={Elsevier}
}

@inproceedings{holder2021efficient,
  title={Efficient uncertainty estimation in semantic segmentation via distillation},
  author={Holder, Christopher J and Shafique, Muhammad},
  booktitle={Proceedings of the {IEEE/CVF} International Conference on Computer Vision},
  pages={3087--3094},
  year={2021}
}

@incollection{salahuddin2023leveraging,
  title={Leveraging Uncertainty Estimation for Segmentation of Kidney, Kidney Tumor and Kidney Cysts},
  author={Salahuddin, Zohaib and Kuang, Sheng and Lambin, Philippe and Woodruff, Henry C},
  booktitle={International Challenge on Kidney and Kidney Tumor Segmentation},
  pages={40--46},
  year={2023},
  publisher={Springer}
}

@incollection{michaud2023using,
  title={Using Uncertainty Information for Kidney Tumor Segmentation},
  author={Michaud, Joffrey and Arega, Tewodros Weldebirhan and Bricq, Stephanie},
  booktitle={International Challenge on Kidney and Kidney Tumor Segmentation},
  pages={54--59},
  year={2023},
  publisher={Springer}
}

@article{nnunet,
  title={{nnU-Net}: a self-configuring method for deep learning-based biomedical image segmentation},
  author={Isensee, Fabian and Jaeger, Paul F and Kohl, Simon AA and Petersen, Jens and Maier-Hein, Klaus H},
  journal={Nature Methods},
  volume={18},
  number={2},
  pages={203--211},
  year={2021},
  publisher={Nature Publishing Group}
}

@article{causey2021ensemble,
  title={An ensemble of {U-Net} models for kidney tumor segmentation with {CT} images},
  author={Causey, Jason and Stubblefield, Jonathan and Qualls, Jake and Fowler, Jennifer and Cai, Lingrui and Walker, Karl and Guan, Yuanfang and Huang, Xiuzhen},
  journal={IEEE/ACM Transactions on Computational Biology and Bioinformatics},
  volume={19},
  number={3},
  pages={1387--1392},
  year={2021},
  publisher={IEEE}
}

@article{santini2019kidney,
  title={Kidney tumor segmentation using an ensembling multi-stage deep learning approach. A contribution to the {KiTS19} challenge},
  author={Santini, Gianmarco and Moreau, No{\'e}mie and Rubeaux, Mathieu},
  journal={arXiv preprint arXiv:1909.00735},
  year={2019}
}

@article{wang2019aleatoric,
  title={Aleatoric uncertainty estimation with test-time augmentation for medical image segmentation with convolutional neural networks},
  author={Wang, Guotai and Li, Wenqi and Aertsen, Michael and Deprest, Jan and Ourselin, S{\'e}bastien and Vercauteren, Tom},
  journal={Neurocomputing},
  volume={338},
  pages={34--45},
  year={2019},
  publisher={Elsevier}
}

@misc{heller2023kits21,
      title={The {KiTS21} Challenge: Automatic segmentation of kidneys, renal tumors, and renal cysts in corticomedullary-phase CT}, 
      author={Nicholas Heller and Fabian Isensee and Dasha Trofimova and Resha Tejpaul and Zhongchen Zhao and Huai Chen and Lisheng Wang and Alex Golts and Daniel Khapun and Daniel Shats and Yoel Shoshan and Flora Gilboa-Solomon and Yasmeen George and Xi Yang and Jianpeng Zhang and Jing Zhang and Yong Xia and Mengran Wu and Zhiyang Liu and Ed Walczak and Sean McSweeney and Ranveer Vasdev and Chris Hornung and Rafat Solaiman and Jamee Schoephoerster and Bailey Abernathy and David Wu and Safa Abdulkadir and Ben Byun and Justice Spriggs and Griffin Struyk and Alexandra Austin and Ben Simpson and Michael Hagstrom and Sierra Virnig and John French and Nitin Venkatesh and Sarah Chan and Keenan Moore and Anna Jacobsen and Susan Austin and Mark Austin and Subodh Regmi and Nikolaos Papanikolopoulos and Christopher Weight},
      year={2023},
      eprint={2307.01984},
      archivePrefix={arXiv},
      primaryClass={cs.CV}
}

@inproceedings{smith2017cyclical,
  title={Cyclical learning rates for training neural networks},
  author={Smith, Leslie N},
  booktitle={2017 IEEE Winter Conference on Applications of Computer Vision (WACV)},
  pages={464--472},
  year={2017},
  organization={IEEE}
}

@article{heller2021state,
  title={The state of the art in kidney and kidney tumor segmentation in contrast-enhanced CT imaging: Results of the KiTS19 challenge},
  author={Heller, Nicholas and Isensee, Fabian and Maier-Hein, Klaus H and Hou, Xiaoshuai and Xie, Chunmei and Li, Fengyi and Nan, Yang and Mu, Guangrui and Lin, Zhiyong and Han, Miofei and others},
  journal={Medical Image Analysis},
  volume={67},
  pages={101821},
  year={2021},
  publisher={Elsevier}
}

@article{yogananda2020fully,
  title={A fully automated deep learning network for brain tumor segmentation},
  author={Yogananda, Chandan Ganesh Bangalore and Shah, Bhavya R and Vejdani-Jahromi, Maryam and Nalawade, Sahil S and Murugesan, Gowtham K and Yu, Frank F and Pinho, Marco C and Wagner, Benjamin C and Emblem, Kyrre E and Bj{\o}rnerud, Atle and others},
  journal={Tomography},
  volume={6},
  number={2},
  pages={186--193},
  year={2020},
  publisher={Multidisciplinary Digital Publishing Institute}
}

@incollection{myronenko2023automated,
  title={Automated {3D} Segmentation of Kidneys and Tumors in {MICCAI KiTS} 2023 Challenge},
  author={Myronenko, Andriy and Yang, Dong and He, Yufan and Xu, Daguang},
  booktitle={International Challenge on Kidney and Kidney Tumor Segmentation},
  pages={1--7},
  year={2023},
  publisher={Springer}
}

@incollection{uhm2023exploring,
  title={Exploring {3D U-Net} Training Configurations and Post-processing Strategies for the {MICCAI} 2023 Kidney and Tumor Segmentation Challenge},
  author={Uhm, Kwang-Hyun and Cho, Hyunjun and Xu, Zhixin and Lim, Seohoon and Jung, Seung-Won and Hong, Sung-Hoo and Ko, Sung-Jea},
  booktitle={International Challenge on Kidney and Kidney Tumor Segmentation},
  pages={8--13},
  year={2023},
  publisher={Springer}
}

@incollection{liu2023dynamic,
  title={Dynamic Resolution Network for Kidney Tumor Segmentation},
  author={Liu, Shuolin and Han, Bing},
  booktitle={International Challenge on Kidney and Kidney Tumor Segmentation},
  pages={14--21},
  year={2023},
  publisher={Springer}
}

@incollection{stoica2023analyzing,
  title={Analyzing domain shift when using additional data for the {MICCAI KiTS23} Challenge},
  author={Stoica, George and Breaban, Mihaela and Barbu, Vlad},
  booktitle={International Challenge on Kidney and Kidney Tumor Segmentation},
  pages={22--29},
  year={2023},
  publisher={Springer}
}

@incollection{qian2023hybrid,
  title={A Hybrid Network Based on {nnU-Net} and Swin Transformer for Kidney Tumor Segmentation},
  author={Qian, Lifei and Luo, Ling and Zhong, Yuanhong and Zhong, Daidi},
  booktitle={International Challenge on Kidney and Kidney Tumor Segmentation},
  pages={30--39},
  year={2023},
  publisher={Springer}
}

@article{loshchilov2016sgdr,
  title={{SGDR}: Stochastic gradient descent with warm restarts},
  author={Loshchilov, Ilya and Hutter, Frank},
  journal={arXiv preprint arXiv:1608.03983},
  year={2016}
}

@inproceedings{baumgartner2019phiseg,
  title={{PHiSeg}: Capturing uncertainty in medical image segmentation},
  author={Baumgartner, Christian F and Tezcan, Kerem C and Chaitanya, Krishna and H{\"o}tker, Andreas M and Muehlematter, Urs J and Schawkat, Khoschy and Becker, Anton S and Donati, Olivio and Konukoglu, Ender},
  booktitle={Medical Image Computing and Computer Assisted Intervention--MICCAI 2019: 22nd International Conference, Shenzhen, China, October 13--17, 2019, Proceedings, Part II 22},
  pages={119--127},
  year={2019},
  organization={Springer}
}

@inproceedings{zhao2022efficient,
  title={Efficient Bayesian uncertainty estimation for nnU-Net},
  author={Zhao, Yidong and Yang, Changchun and Schweidtmann, Artur and Tao, Qian},
  booktitle={International Conference on Medical Image Computing and Computer-Assisted Intervention},
  pages={535--544},
  year={2022},
  organization={Springer}
}

@article{khalili2024uncertainty,
  title={Uncertainty-guided annotation enhances segmentation with the human-in-the-loop},
  author={Khalili, Nadieh and Spronck, Joey and Ciompi, Francesco and van der Laak, Jeroen and Litjens, Geert},
  journal={arXiv preprint arXiv:2404.07208},
  year={2024}
}

@article{mehrtash2020confidence,
  title={Confidence calibration and predictive uncertainty estimation for deep medical image segmentation},
  author={Mehrtash, Alireza and Wells, William M and Tempany, Clare M and Abolmaesumi, Purang and Kapur, Tina},
  journal={IEEE Transactions on Medical Imaging},
  volume={39},
  number={12},
  pages={3868--3878},
  year={2020},
  publisher={IEEE}
}

@article{monteiro2020stochastic,
  title={Stochastic segmentation networks: Modelling spatially correlated aleatoric uncertainty},
  author={Monteiro, Miguel and Le Folgoc, Lo{\"\i}c and Coelho de Castro, Daniel and Pawlowski, Nick and Marques, Bernardo and Kamnitsas, Konstantinos and van der Wilk, Mark and Glocker, Ben},
  journal={Advances in Neural Information Processing Systems},
  volume={33},
  pages={12756--12767},
  year={2020}
}

@article{saman2019survey,
  title={Survey on brain tumor segmentation and feature extraction of MR images},
  author={Saman, Sangeetha and Jamjala Narayanan, Swathi},
  journal={International journal of multimedia information retrieval},
  volume={8},
  pages={79--99},
  year={2019},
  publisher={Springer}
}

@InProceedings{SDV21,
  author = {Sakaridis, Christos and Dai, Dengxin and Van Gool, Luc},
  title = {{ACDC}: The Adverse Conditions Dataset with Correspondences for Semantic Driving Scene Understanding},
  booktitle = {Proceedings of the IEEE/CVF International Conference on Computer Vision (ICCV)},
  month = {October},
  year = 2021
}

@article{campello2021multi,
  title={Multi-centre, multi-vendor and multi-disease cardiac segmentation: the M\&Ms challenge},
  author={Campello, Victor M and Gkontra, Polyxeni and Izquierdo, Cristian and Martin-Isla, Carlos and Sojoudi, Alireza and Full, Peter M and Maier-Hein, Klaus and Zhang, Yao and He, Zhiqiang and Ma, Jun and others},
  journal={IEEE Transactions on Medical Imaging},
  volume={40},
  number={12},
  pages={3543--3554},
  year={2021},
  publisher={IEEE}
}

@misc{code,
  author       = {Sina Ziaee},
  title        = {{Rel-UNet: Reliable Tumor Segmentation via Uncertainty Quantification in nnU-Net}},
  year         = {2024},
  url          = {https://github.com/sinaziaee/sgdr},
  note         = {GitHub repository}
}

\begin{thebibliography}{00}

\bibitem{jungo2019assessing}
A. Jungo and M. Reyes, 
``Assessing reliability and challenges of uncertainty estimations for medical image segmentation,''
in \emph{Medical Image Computing and Computer Assisted Intervention--MICCAI 2019: 22nd International Conference, Shenzhen, China, October 13--17, 2019, Proceedings, Part II 22}, 
Springer, 2019, pp. 48--56.

\bibitem{franchi2021one}
G. Franchi, A. Bursuc, E. Aldea, S. Dubuisson, and I. Bloch, 
``One versus all for deep neural network for uncertainty (OVNNI) quantification,''
\emph{IEEE Access}, vol. 10, pp. 7300--7312, 2021.

\bibitem{roy2022uncertainty}
S. Roy, M. Trapp, A. Pilzer, J. Kannala, N. Sebe, E. Ricci, and A. Solin,
``Uncertainty-guided source-free domain adaptation,''
in \emph{European Conference on Computer Vision}, 
Springer, 2022, pp. 537--555.

\bibitem{dorta2018structured}
G. Dorta, S. Vicente, L. Agapito, N. D. F. Campbell, and I. Simpson, 
``Structured uncertainty prediction networks,''
in \emph{Proc. IEEE Conf. on Computer Vision and Pattern Recognition (CVPR)}, 
2018, pp. 5477--5485.

\bibitem{zou2023review}
K. Zou, Z. Chen, X. Yuan, X. Shen, M. Wang, and H. Fu,
``A review of uncertainty estimation and its application in medical imaging,''
\emph{Meta-Radiology}, p. 100003, 2023.

\bibitem{holder2021efficient}
C. J. Holder and M. Shafique,
``Efficient uncertainty estimation in semantic segmentation via distillation,''
in \emph{Proc. IEEE/CVF Int. Conf. on Computer Vision (ICCV)}, 
2021, pp. 3087--3094.

\bibitem{salahuddin2023leveraging}
Z. Salahuddin, S. Kuang, P. Lambin, and H. C. Woodruff,
``Leveraging Uncertainty Estimation for Segmentation of Kidney, Kidney Tumor and Kidney Cysts,''
in \emph{International Challenge on Kidney and Kidney Tumor Segmentation}, 
Springer, 2023, pp. 40--46.

\bibitem{michaud2023using}
J. Michaud, T. W. Arega, and S. Bricq,
``Using Uncertainty Information for Kidney Tumor Segmentation,''
in \emph{International Challenge on Kidney and Kidney Tumor Segmentation}, 
Springer, 2023, pp. 54--59.

\bibitem{nnunet}
F. Isensee, P. F. Jaeger, S. A. A. Kohl, J. Petersen, and K. H. Maier-Hein,
``nnU-Net: a self-configuring method for deep learning-based biomedical image segmentation,''
\emph{Nature Methods}, vol. 18, no. 2, pp. 203--211, 2021.

\bibitem{causey2021ensemble}
J. Causey, J. Stubblefield, J. Qualls, J. Fowler, L. Cai, K. Walker, Y. Guan, and X. Huang,
``An ensemble of U-Net models for kidney tumor segmentation with CT images,''
\emph{IEEE/ACM Trans. on Computational Biology and Bioinformatics}, vol. 19, no. 3, pp. 1387--1392, 2021.

\bibitem{santini2019kidney}
G. Santini, N. Moreau, and M. Rubeaux,
``Kidney tumor segmentation using an ensembling multi-stage deep learning approach. A contribution to the KiTS19 challenge,''
\emph{arXiv preprint arXiv:1909.00735}, 2019.

\bibitem{wang2019aleatoric}
G. Wang, W. Li, M. Aertsen, J. Deprest, S. Ourselin, and T. Vercauteren,
``Aleatoric uncertainty estimation with test-time augmentation for medical image segmentation with convolutional neural networks,''
\emph{Neurocomputing}, vol. 338, pp. 34--45, 2019.

\bibitem{heller2023kits21}
N. Heller, F. Isensee, D. Trofimova \emph{et al.},
``The KiTS21 Challenge: Automatic segmentation of kidneys, renal tumors, and renal cysts in corticomedullary-phase CT,''
\emph{arXiv preprint arXiv:2307.01984}, 2023.

\bibitem{smith2017cyclical}
L. N. Smith,
``Cyclical learning rates for training neural networks,''
in \emph{Proc. IEEE Winter Conf. on Applications of Computer Vision (WACV)},
2017, pp. 464--472.

\bibitem{heller2021state}
N. Heller, F. Isensee, K. H. Maier-Hein \emph{et al.},
``The state of the art in kidney and kidney tumor segmentation in contrast-enhanced CT imaging: Results of the KiTS19 challenge,''
\emph{Medical Image Analysis}, vol. 67, p. 101821, 2021.

\bibitem{yogananda2020fully}
C. G. B. Yogananda, B. R. Shah, M. Vejdani-Jahromi \emph{et al.},
``A fully automated deep learning network for brain tumor segmentation,''
\emph{Tomography}, vol. 6, no. 2, pp. 186--193, 2020.

\bibitem{myronenko2023automated}
A. Myronenko, D. Yang, Y. He, and D. Xu,
``Automated 3D Segmentation of Kidneys and Tumors in MICCAI KiTS 2023 Challenge,''
in \emph{International Challenge on Kidney and Kidney Tumor Segmentation}, 
Springer, 2023, pp. 1--7.

\bibitem{uhm2023exploring}
K.-H. Uhm, H. Cho, Z. Xu \emph{et al.},
``Exploring 3D U-Net Training Configurations and Post-processing Strategies for the MICCAI 2023 Kidney and Tumor Segmentation Challenge,''
in \emph{International Challenge on Kidney and Kidney Tumor Segmentation}, 
Springer, 2023, pp. 8--13.

\bibitem{liu2023dynamic}
S. Liu and B. Han,
``Dynamic Resolution Network for Kidney Tumor Segmentation,''
in \emph{International Challenge on Kidney and Kidney Tumor Segmentation}, 
Springer, 2023, pp. 14--21.

\bibitem{stoica2023analyzing}
G. Stoica, M. Breaban, and V. Barbu,
``Analyzing domain shift when using additional data for the MICCAI KiTS23 Challenge,''
in \emph{International Challenge on Kidney and Kidney Tumor Segmentation}, 
Springer, 2023, pp. 22--29.

\bibitem{qian2023hybrid}
L. Qian, L. Luo, Y. Zhong, and D. Zhong,
``A Hybrid Network Based on nnU-Net and Swin Transformer for Kidney Tumor Segmentation,''
in \emph{International Challenge on Kidney and Kidney Tumor Segmentation}, 
Springer, 2023, pp. 30--39.

\bibitem{loshchilov2016sgdr}
I. Loshchilov and F. Hutter,
``SGDR: Stochastic gradient descent with warm restarts,''
\emph{arXiv preprint arXiv:1608.03983}, 2016.

\bibitem{baumgartner2019phiseg}
C. F. Baumgartner, K. C. Tezcan, K. Chaitanya \emph{et al.},
``PHiSeg: Capturing uncertainty in medical image segmentation,''
in \emph{Medical Image Computing and Computer Assisted Intervention--MICCAI 2019: 22nd International Conference, Shenzhen, China, October 13--17, 2019, Proceedings, Part II 22},
Springer, 2019, pp. 119--127.

\bibitem{zhao2022efficient}
Y. Zhao, C. Yang, A. Schweidtmann, and Q. Tao,
``Efficient Bayesian uncertainty estimation for nnU-Net,''
in \emph{International Conference on Medical Image Computing and Computer-Assisted Intervention (MICCAI)},
Springer, 2022, pp. 535--544.

\bibitem{khalili2024uncertainty}
N. Khalili, J. Spronck, F. Ciompi, J. van der Laak, and G. Litjens,
``Uncertainty-guided annotation enhances segmentation with the human-in-the-loop,''
\emph{arXiv preprint arXiv:2404.07208}, 2024.

\bibitem{mehrtash2020confidence}
A. Mehrtash, W. M. Wells, C. M. Tempany, P. Abolmaesumi, and T. Kapur,
``Confidence calibration and predictive uncertainty estimation for deep medical image segmentation,''
\emph{IEEE Trans. on Medical Imaging}, vol. 39, no. 12, pp. 3868--3878, 2020.

\bibitem{monteiro2020stochastic}
M. Monteiro, L. Le Folgoc, D. C. de Castro \emph{et al.},
``Stochastic segmentation networks: Modelling spatially correlated aleatoric uncertainty,''
\emph{Advances in Neural Information Processing Systems}, vol. 33, pp. 12756--12767, 2020.

\bibitem{saman2019survey}
S. Saman and S. J. Narayanan,
``Survey on brain tumor segmentation and feature extraction of MR images,''
\emph{International Journal of Multimedia Information Retrieval}, vol. 8, pp. 79--99, 2019.

\bibitem{SDV21}
C. Sakaridis, D. Dai, and L. Van Gool,
``ACDC: The Adverse Conditions Dataset with Correspondences for Semantic Driving Scene Understanding,''
in \emph{Proc. IEEE/CVF Int. Conf. on Computer Vision (ICCV)}, Oct. 2021.

\bibitem{campello2021multi}
V. M. Campello, P. Gkontra, C. Izquierdo \emph{et al.},
``Multi-centre, multi-vendor and multi-disease cardiac segmentation: the M\&Ms challenge,''
\emph{IEEE Trans. on Medical Imaging}, vol. 40, no. 12, pp. 3543--3554, 2021.

\bibitem{code}
S. Ziaee,
\emph{Rel-UNet: Reliable Tumor Segmentation via Uncertainty Quantification in nnU-Net}, 
GitHub repository, 2024. [Online]. Available: \url{https://github.com/sinaziaee/sgdr}

\end{thebibliography}

\appendix

\section{Supplementary Material}

\subsection{Experimental Setup}
The training is performed using nnU-Net ResEnc Large model, SGD as optimizer, SGDR as learning rate scheduler, initial learning rate with 0.1, T\_0 with 100 (The number of epochs before the first restart), T\_mult with 2 (The factor by which the cycle length (T\_0) is multiplied after each restart), eta\_min is 0.0001 (minimum learning rate). The number of epochs is 800. Thus, with the configuration above, the restart happens around epochs 100, 200, 400, and 800. The loss function is Dice Loss, and the batch size is 40.

\subsection{Comparison with other experiments}
We ran different setups of our method to compare the performance of segmentation and uncertainty maps under different variations. Table~\ref{tab:my_way_comparison} shows the comparison of 5-fold, 3D, and Full train comparison. 5-fold is training five different models with five different folds of the dataset and create an ensemble of models for uncertainty generation. 3D is the exact configuration of our method but with the original 3D images. Full-train is training a single model and using the epochs at 60\%, 70\%, 80\%, 90\%, and the last epoch to create the ensemble of checkpoint models and generate uncertainty maps. Analysis of Table~\ref{tab:my_way_comparison} is as follows:
\begin{itemize}
    \item 5-fold approach has a higher Dice score and lower entropy compared to the SGDR approach that we used for training, but it also has higher ECE. This means the model could be achieving a high Dice score by correctly predicting most of the correct segmentations, but when it makes mistakes, it does so with high confidence, as it is not well-calibrated.
    \item Full-train setup indicates low ECE, which means it does not show good calibration and has high entropy (uncertainty).
    \item 3D experiment with the same 800 training epochs shows a lower Dice score, higher ECE, and higher entropy. Although 3D segmentation considers the voxels in its predictions, it does not necessarily perform better as we see in this experiment. The reason is that with the same architecture and the same number of epochs for training, it does not perform well, and 
    the model requires a deeper architecture with more epochs to perform well. In addition, training in 3D architecture requires more training data than 2D training in the same architecture. With 3D, we have at most 458 cases and images in KiTS23, but with 2D, we have more than 90000 images to train after the 2D slicing of the CT scan images.
\end{itemize}

\begin{figure}
    \centering
    \includegraphics[width=1.0\linewidth]{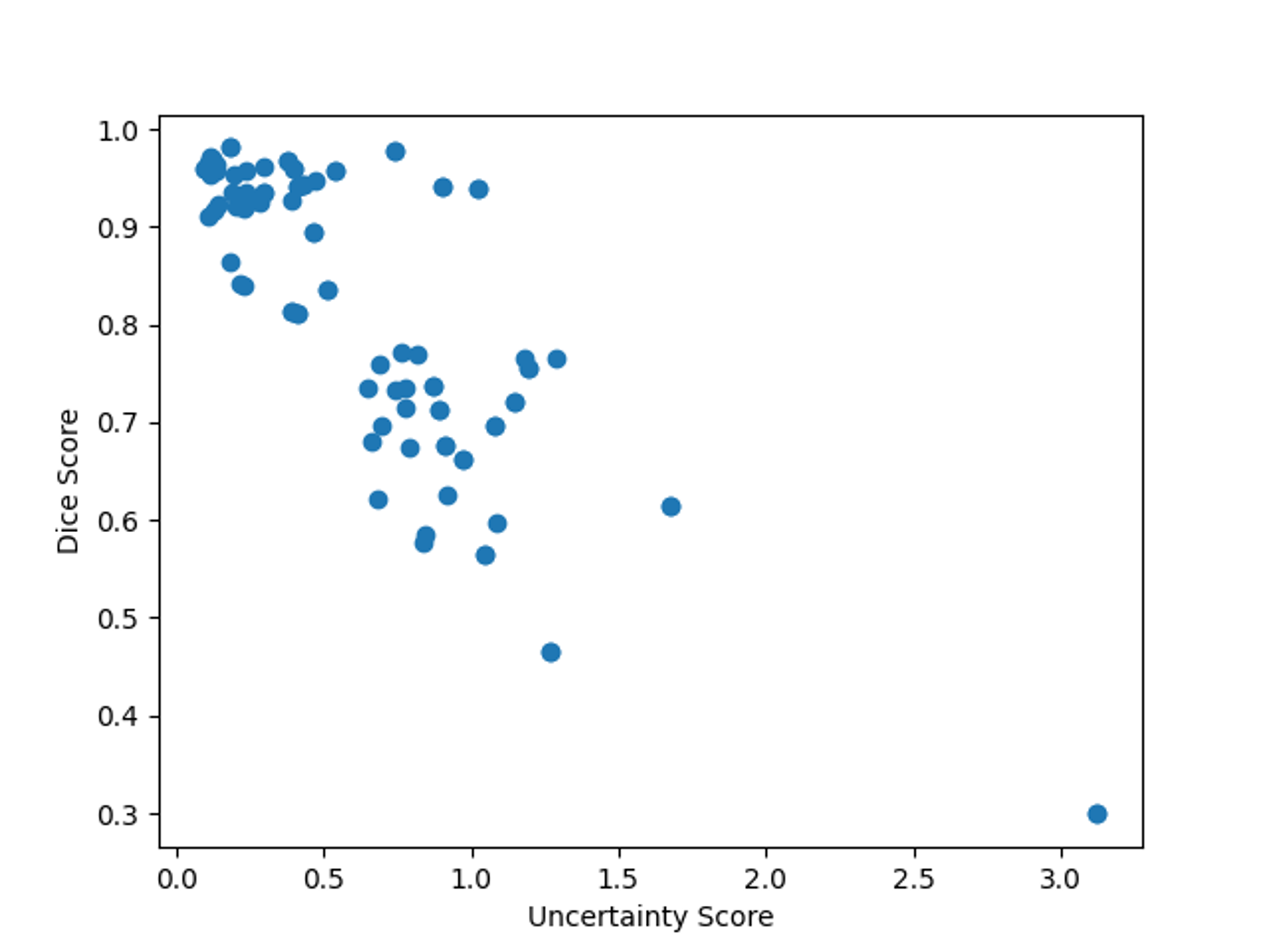}
    \caption{Scatterplot showing the relationship between Uncertainty (x-axis) and Dice score (y-axis). The graph indicates that lower uncertainty usually concludes higher segmentation performance. It indicates that segmentation performance (Dice score) is more than 80\% when the uncertainty score is less than 0.5.}
    \label{fig:dice_vs_uncertainty}
\end{figure}

\begin{table}[t]
\caption{Comparison of our approach with 5-fold cross-validation model, the 3D architecture of our approach, and full-train using ECE and Dice score.}
    \centering
    \setlength{\tabcolsep}{5pt} %
    \begin{tabular}{c|c|c|c}
         Method & ECE (\%) & Dice score (\%) & Avg. Entropy\\
         \hline 
         5-fold & 3.02 & 81.1 & 0.52\\
         3D & 5.20 & 73.3 & 0.98\\
         Full-train & 3.37 & 80.5 & 0.71\\
         \textbf{Rel-UNet} & \textbf{1.11} & \textbf{80.0} & \textbf{0.67}\\
    \end{tabular}
    \label{tab:my_way_comparison}
\end{table}

\begin{table}[ht]
\caption{Comparison of our SGDR segmentation model with top 6 performers in KITS23 challenge. The purpose of our model is not segmentation alone, and we incorporate uncertainty in our prediction in contrast to these top 6 performers.}
    \centering
    \setlength{\tabcolsep}{3pt} %
    \begin{tabular}{c|c|c|c}
         Method & Kidney Dice & Masses Dice & Tumor Dice  \\
         \hline 
         Andriy Myronenko et al. \cite{myronenko2023automated} & 0.835 & 0.723 & 0.758 \\
         Kwang-Hyun Uhm et al. \cite{uhm2023exploring} & 0.820 & 0.712 & 0.738 \\
         Yasmeen George et al. & 0.819 & 0.707 & 0.713 \\
         Shuolin Liu et al. \cite{liu2023dynamic} & 0.805 & 0.706 & 	0.697 \\
         George Stoica et al. \cite{stoica2023analyzing} & 0.807 & 0.691 & 0.713 \\
         Lifei Qian et al. \cite{qian2023hybrid} & 0.801 & 0.680 & 0.687 \\
         \textbf{Rel-UNet} & 0.800 & 70.2 & 0.641 \\
    \end{tabular}
    \label{tab:results}
\end{table}


\subsection{Comparison in Segmentation}
Table~\ref{tab:results} indicates the segmentation result compared to the top 6 performers in the KiTS23 challenge. Although the segmentation result of our model is not as good as the result in the top 6 methods, our aim is not to solely perform segmentation but to incorporate uncertainty to qualify the segmentation results more precisely.

\subsection{Uncertainty and Segmentation Performance}
Figure~\ref{fig:dice_vs_uncertainty} shows the segmentation performance and the uncertainty score calculated from the test dataset. With some approximation, the segmentation model performs better in areas with lower uncertainty scores. In cases where the uncertainty score is less than 0.5, the Dice score (segmentation performance) is higher than 80\%.

\end{document}